\documentstyle[preprint,aps]{revtex}

\begin{document}

\title{\bf Mixing of $d_{x^2-y^2}$ and $d_{xy}$ superconducting states 
for different filling and temperature
\thanks{PACS \# 74.20.Fg, 74.62.-c, 74.25.Bt. KEYWORDS
{mixed-symmetry state, high-$T_c$ superconductor, $d_{x^2-y^2}$ and
$d_{xy}$ order parameters}}}


\author{Angsula Ghosh$^1$ and Sadhan K Adhikari$^2$\thanks{Corresponding 
author: e-mail: adhikari@ift.unesp.br Fax 011 55 11 3177 9080 Address
Instituto de Fisica Teorica, Rua Pamplona 145, 01405-900 S\~ao Paulo, SP,
Brazil 
}}

\address{$^1$Instituto de F\'{\i}sica, 
Universidade de S\~ao Paulo,
Caixa Postal 66318, 05315-970 S\~ao Paulo, S\~ao Paulo, Brazil\\
$^2$Instituto de F\'{\i}sica Te\'orica, 
Universidade Estadual Paulista,
01.405-900 S\~ao Paulo, S\~ao Paulo, Brazil}

\maketitle

\begin{abstract}

We investigate the solution of the gap equation for mixed order parameter
symmetry states as a function of filling using a two-dimensional
tight-binding model incorporating second-neighbor hopping for tetragonal
and orthorhombic lattice. The principal (major) component of the order
parameter is taken to be of the $d_{x^2-y^2}$ type. As suggested in
several investigations the minor component of the order parameter is taken
to be of the $d_{xy}$ type. Both the permissible mixing angles 0 and $\pi
/2$ between the two components are considered. As a function of filling
pronounced maxima of $d_{x^2-y^2}$ order parameter is accompanied by
minima of the $d_{xy}$ order parameter. At fixed filling, the temperature
dependence of the two components of the order parameter is also studied in
all cases. The variation of critical temperature $T_c$ with filling is
also studied and $T_c$ is found to increase with second-neighbor hopping.

PACS \# 74.20.Fg, 74.62.-c, 74.25.Bt. 

KEYWORDS
mixed-symmetry state, high-$T_c$ superconductor, order parameter

\end{abstract} 


\newpage

\section{Introduction}

Many years after the discovery of the unconventional high-$T_c$
cuprates \cite{1} with a high critical temperature $T_c$ the
symmetry of its order parameter continues the subject of active
investigation. From both experimental observations 
\cite{2a,2b,3a,3b,3c} and related theoretical
analyses \cite{4a,4b,4c,4d,4e,5a1,5a2,5a3}
there is a consensus that immediately below $T_c$, the symmetry
of the order parameter is of the $d_{x^2-y^2}$ type.
However, many
experiments \cite{5a,5b,5c,5d,6,7,8a,8b,9,10} and theoretical analyses
\cite{11a,11b,131,132,133,134,135,136,13a1,13a2,14a,14b,14c,14d,14e}
suggest that at lower
temperatures
the order parameter has a mixed symmetry of the $d_{x^2-y^2}+\exp
(i\theta)\chi $ type, where $\chi$ represents a minor-component state with
a distinct symmetry superposed on the major component $d_{x^2-y^2}$.  From
theoretical analysis the mixing angle $\theta$ can have the values $0$,
$\pi /2, \pi,$ and $ 3\pi/2$. For mixing angles $\pi/2$ and $3\pi/2$ the
time-reversal symmetry is broken for these states \cite{11a,11b}.
Four possible candidates for the minor
symmetry state $\chi$ are
the $d_{xy}$, $s$, $s_{x^2+y^2}$ and $s_{xy}$ states.  From a group
theoretical point of view these states belong to the same irreducible
representation of the orthorhombic point group.  However, there is still
controversy about the nature of the minor component and the value of the
mixing angle in different high-$T_c$ materials. There have been extensive
discussions in recent literature
about the nature of the minor component and the value of the mixing angle.
Without repeating those discussions in detail here we
present a brief summary of these analyses which suggest a mixed-symmetry
state and refer the interested reader to these references
\cite{11a,11b,131,132,133,134,135,136,14a,14b,14c,14d,14e}.

There are experimental evidences based on the Josephson supercurrent for
tunneling between a conventional $s$-wave superconductor (Pb) and twinned
or untwinned single crystals of YBa$_2$Cu$_3$O$_7$ (YBCO) that YBCO has a
mixed-symmetry state with a minor $s$-wave component at lower temperatures
\cite{5a,5b,5c,5d}.  Kouznetsov {\it et al.} \cite{6} performed  $c$-axis
Josephson
tunneling experiments by depositing a conventional superconductor (Pb)
across the single twin boundary of a YBCO crystal and from a study of the
critical current under an applied magnetic field they also found evidence
of a mixed-symmetry state with a minor $s$-wave component. Similar
conclusion was reached by Sridhar {\it et al.} \cite{7} from the
measurement of the
microwave complex conductivity of high quality YBa$_2$Cu$_3$O$_{7-\delta}$
single crystals at 10 GHz using a high-$Q$ Nb cavity. 

From an analysis of angle-resolved photoemission spectroscopy experiment,
Ma {\it et al.} \cite{8a,8b} detected a temperature-dependent gap
anisotropy
in Bi$_2$Sr$_2$CaCu$_2$O$_{8+x}$ which can be associated with a
mixed-symmetry state with a minor $d_{xy}$ component. More recently, Krishana
{\it et al.} \cite{9} reported a phase transition to a mixed-symmetry
state with a minor $d_{xy}$ component in Bi$_2$Sr$_2$CaCu$_2$O$_{8}$
induced by a magnetic field from a study of thermal conductivity as a
function of temperature and applied field. Later, the experimental study
of thermal conductivity in Bi$_2$Sr$_2$CaCu$_2$O$_{8}$ by Movshovich {\it
et al.} \cite{10} also leads to the same conclusion. 

Different theoretical investigations \cite{11a,11b} provided explanations
of
the observation by Krishana {\it et al. } that for weak magnetic field
possibly a minor time-reversal symmetry-breaking state is induced on a
dominating $d_{x^2-y^2}$ state in Bi$_2$Sr$_2$CaCu$_2$O$_{8}$.
 This mixed-symmetry state is likely to be a minor $d_{xy}$
component on a $d_{x^2-y^2}$ state at lower temperatures.  The above
investigations in Bi$_2$Sr$_2$CaCu$_2$O$_{8}$ suggest a minor $d_{xy}$
component in the order parameter.  In view of this we perform a
theoretical study of this problem using a tight-binding model including 
orthorhombic distortion and second-nearest-neighbor hopping.  Both
orthorhombicity and filling are expected to play a crucial role in the
evolution of these mixed-symmetry states and we study in this paper the
properties of
the mixed-symmetry states of the type $d_{x^2-y^2}+\chi d_{xy}$ at
different temperature, filling, orthorhombicity, and second-neighbor
hopping. A similar study of the mixture of the $d_{x^2-y^2}$ and different 
types of $s$ states will be reported elsewhere.

There is no suitable microscopic theory for high-$T_c$ cuprates and there
is controversy about a proper description of the normal state and the
pairing mechanism for such materials. Also, for underdoped systems the
existence of a pseudo gap above the superconducting critical temperature
can not yet be understood within a microscopic theory. However, with the
increase of doping the superconducting critical temperature increases and
the pseudo gap is reduced and eventually it disappears at optimal doping.
In the absence of a microscopic theory, we use a phenomenological
two-dimensional tight-binding model with appropriate lattice symmetry for
studying some of the general features of the mixed-symmetry states of
overdoped high-$T_c$ superconductors where there is no pseudo gap. This
model has been used
successfully in describing many properties of high-$T_c$ materials
\cite{11a,11b,131,132,133,134,135,136,13a1,13a2}.

In high-$T_c$ cuprates, it is generally believed that superconductivity
resides mainly in the CuO$_2$ planes with nearly tetragonal symmetry.
Hence, quite expectedly, the values of critical temperature and
superconducting gap of high-$T_c$ materials are very sensitive to the
level of doping which determines the number of electrons available for
conduction in the two-dimensional CuO$_2$ plane of the cuprates.  
However, 
a large anisotropy in the penetration depth between the $a$
and $b$ directions has been observed experimentally \cite{ba} in YBCO and
this cannot be understood in a purely tetragonal system.
Therefore,
in YBCO there is the evidence of a weak orthorhombic distortion.  
 Hence we shall
consider orthorhombic distortion in the present study of mixed-symmetry
states.

One of the objectives of the present study is to investigate
how the superconductivity in the uncoupled and mixed-symmetry states
changes with doping or filling.  The present single-band model can
acomodate 
a maximum filling of two electrons (spin up and down) per unit cell.
 In addition to studying the evolution of
the mixed-symmetry states with respect to filling or doping in presence
of second-nearest-neighbor hopping for tetragonal and orthorhombic
lattice, we also study
the temperature dependence of the order parameters for a fixed filling.

In Sec. II we describe the formalism. In Sec. III we present our numerical
results and finally in Sec. IV we give a brief summary. 

\section{Theoretical Formulation}

We base the present study on the two-dimensional  tight binding model 
which we describe below. This model is sufficiently general for considering 
mixed angular momentum states on tetragonal and orthorhombic lattice, 
employing nearest and second-nearest-neighbor hopping integrals.  
In the present tight-binding model  the effective interaction $V_{\bf k
q}$ for transition from a momentum $\bf q$ to $\bf k$ is taken to be
separable, and is expanded in terms of some general basis functions $\eta
_{i\bf k}$, labeled by index $i$, so that 
\begin{equation}\label{1}
V_{\bf k q}=-\sum_i V_i \eta _{i\bf k}\eta _{i\bf q}
\end{equation}  
The separable nature of the interaction facilitates the solution of the
gap equation.
The orthogonal functions $\eta _{i\bf k}$ are associated with a 
one-dimensional irreducible representation of the point group of square
lattice $C_{4v}$ and are appropriate generalizations of the circular
harmonics incorporating the proper lattice symmetry. Here $V_i$ is the
coupling of effective interaction in the specific angular momentum state
specified by the index $i$ above. In general we have the following
$\eta$'s
\begin{eqnarray}
\eta _{1\bf q}& =& \cos q_x -\beta \cos q_y, \hskip 1cm  d_{x^2-y^2}
\mbox{-wave},\\    
    \eta _{2\bf q}&= &2\sin q_x \sin q_y,  \hskip 1.5cm d_{xy}
\mbox{-wave},
\\
\eta _{3\bf q}&=& 1,  \hskip 3.2cm s \mbox{-wave}, \\
\eta _{4\bf q}&= &\cos q_x +\beta \cos q_y,  \hskip 1cm s_{x^2+y^2}
\mbox{-wave},\\    
    \eta _{5\bf q}&= &2\cos q_x \cos q_y,  \hskip 1.34cm  s_{xy}
\mbox{-wave},  
\end{eqnarray} 
etc. 
Here $\beta  =1$ corresponds to square lattice and $\beta \ne 1$
represents orthorhombic distortion.  
The orthogonality property of functions $\eta$'s  is given by
\begin{eqnarray}\label{2}
\sum_{\bf q} \eta _{i\bf q}\eta _{j\bf q} = 0, ..., i\ne j.
\end{eqnarray} 

In the present investigation of mixed-symmetry states involving
$d_{x^2-y^2 }$ and $d_{xy}$, it is appropriate to consider only two terms
in the sum (\ref{1}) with $\eta_1$ and $\eta_2$ above.  We shall consider
singlet Cooper pairing and subsequent  condensation in $d_{x^2-y^2}$
and $d_{xy}$ waves, denoted by indices 1 and 2, respectively, and the
mixed-symmetry state formed by these two. Consequently, the potential
couplings $V_1$ and $V_2$ will be taken to be positive corresponding to
attractive interactions. 

We consider a single tight-binding two-dimensional band with electron
dispersion relation including second-nearest-neighbor hopping. 
In this case the quasiparticle dispersion relation relating the electronic
energy $\epsilon_{\bf k}$  and momentum ${\bf k}$ is taken as 
\begin{equation}\label{3}
\epsilon_{\bf k}=-2t(\cos k_x+\beta \cos k_y-2\gamma\cos k_x
\cos k_y) -\mu,
\end{equation}
where $t$ and $\beta t$ are the nearest-neighbor hopping integrals
along the in-plane $a$ and $b$ axes, respectively, and $\gamma t$ is the 
second-nearest-neighbor hopping integral. In Eq. (\ref{3}) $\mu$ is the
chemical potential measured with respect to the Fermi energy and is
determined once the filling $n$ is specified.  The
nearest-neighbor hopping parameter $t$ is typically taken to be $\sim 0.1$
eV. The parameter $\beta$ destroys the symmetry between the $a$ and $b$
directions in the CuO$_2$ planes in this simple model. The potential $V_{\bf k
q}$ above
also possesses such a symmetry-breaking term. 
The
energy $\epsilon_{\bf k}$ is measured with respect to the Fermi surface.
Such a one-band model with different first-neighbor-hopping parameters in
the $a$ and $b$ directions is the simplest approximate way of including in
the theoretical description the effect of orthorhombicity.

At a finite temperature $T$, one has  the following gap equation
\begin{eqnarray}
\Delta_{\bf k}& =& -\sum_{\bf q} V_{\bf kq}\frac{\Delta_{ \bf q}}{2E_{\bf
q}}\tanh
\frac{E_{\bf q} }{2k_BT}  \label{4} \end{eqnarray} 
with $E_{\bf q} =
[(\epsilon_{\bf q}-\mu  )^2 + |\Delta_{\bf q}|^{2}]^ {1/2},$ 
 and $k_B$ the Boltzmann constant. 

It has been observed that the critical temperature $T_c$ is sensitive to
the level of doping which determines the number of available conduction
electrons in the CuO$_2$ plane. In this model  
the chemical potential $\mu$ and the  the filling $n$ are 
determined by the number equation 
\begin{equation}\label{5}
n= 1- \sum_{\bf q}\frac {\epsilon_{\bf q}-\mu}{E_{\bf
q}}\tanh
\frac{E_{\bf q} }{2k_BT}.
\end{equation} 
The filling $n$ can be related to the the experimental doping $\delta$ in
the three dimensional Brillouin zone by $n=1-\delta$. In this work we
study the  variation of $n$ from 0 to 1 (half filling).

The order parameter $\Delta _{\bf q}$ has the
following anisotropic form: 
\begin{equation}\label{6}
\Delta _{\bf q} =  \Delta_1 \eta_{1\bf q} +C \Delta_2 \eta_{2\bf q},
\end{equation}
where $C\equiv \exp (i\theta) $ is a complex
number of unit modulus $|C|^2 = \cos^2\theta+\sin^2 \theta =1$.  
We substitute Eqs.  (\ref{1}) 
and (\ref{6}) into  the gap equation (\ref{4}) and using the orthogonality 
property (\ref{2}) obtain the two following coupled equations for
$\Delta_1$ and $\Delta_2$: 
\begin{eqnarray}\label{7}
\Delta_1 &=& \sum_{\bf q}V_1\eta_{1\bf q}\frac {\Delta_1 \eta_{1\bf q}
+C \Delta_2 \eta_{2\bf q}}{2E_{\bf q}}\tanh \frac{E_{\bf q}}{2k_BT},\\
C\Delta_2 &=& \sum_{\bf q}V_2\eta_{2\bf q}\frac {\Delta_1 \eta_{1\bf q}
+C \Delta_2 \eta_{2\bf q}}{2E_{\bf q}}\tanh \frac{E_{\bf q}}{2k_BT}.
\label{8}
\end{eqnarray}

In order to proceed further we have to know the nature of $C$, e. g., 
whether $C$ is real, imaginary or complex. As $|C|^2=1$, a purely real $C$
implies $C=\pm 1$ and in this case no further simplification of the
coupled Eqs. (\ref{7}) and (\ref{8}) is possible and one has 
\begin{eqnarray}\label{7a}
\Delta_1 &=& \sum_{\bf q}V_1\eta_{1\bf q}\frac {\Delta_1 \eta_{1\bf q}
\pm \Delta_2 \eta_{2\bf q}}{2E_{\bf q}}\tanh \frac{E_{\bf q}}{2k_BT},\\
\pm\Delta_2 &=& \sum_{\bf q}V_2\eta_{2\bf q}\frac {\Delta_1 \eta_{1\bf q}
\pm \Delta_2 \eta_{2\bf q}}{2E_{\bf q}}\tanh \frac{E_{\bf q}}{2k_BT},
\label{8a}
\end{eqnarray}
with 
\begin{equation}\label{7aa}
E_{\bf q} =
[(\epsilon_{\bf q}-\mu  )^2 + (\Delta_1 \eta_{1\bf q} \pm \Delta_2
\eta_{2\bf q})^2]^ {1/2},
\end{equation}
The upper sign in Eqs. (\ref{7a}), (\ref{8a}), and (\ref{7aa})
corresponds to a mixing
angle $\theta
=0$ ($C=1$)  and the lower sign to $\theta=\pi$ ($C=-1$). 

 However, Eqs.
(\ref{7}) and (\ref{8}) get
simplified when $C$ is imaginary, e. g., $C=\pm i$. Consequently,  
\begin{equation}\label{9}
E_{\bf q} =
[(\epsilon_{\bf q}-\mu  )^2 + \Delta_1^2 \eta^2_{1\bf q} + \Delta_2^2
\eta^2_{2\bf q}]^ {1/2},
\end{equation}
and Eqs. (\ref{7})  and (\ref{8})  reduce to 
\begin{eqnarray}\label{10}
\Delta_1 &=& \sum_{\bf q}V_1\frac {\Delta_1 \eta^2_{1\bf q}
}{2E_{\bf q}}\tanh \frac{E_{\bf q}}{2k_BT},\\
\Delta_2 &=& \sum_{\bf q}V_2\frac {
 \Delta_2 \eta^2_{2\bf q}}{2E_{\bf q}}\tanh \frac{E_{\bf q}}{2k_BT},
\label{11}
\end{eqnarray}
since in this case 
\begin{eqnarray}\label{12}
 \sum_{\bf q} \frac { \eta_{1\bf q}\eta_{2\bf q}
}{2E_{\bf q}}\tanh \frac{E_{\bf q}}{2k_BT}=0,
\end{eqnarray}
which follows from the definitions of $\eta_{1\bf q}$ and $\eta_{2\bf q}$
and Eq.  (\ref{9}). However, Eq. (\ref{12}), which is responsible for the
simplification for $C=\pm i$ ($\theta= \pi/2$ and $3\pi/2$), does not hold
for $C=\pm 1$ or for a general
complex $C$.

Finally, for a general complex $C$ one can separate in a
straightforward fashion Eqs.  (\ref{7}) and
(\ref{8}) into their real and imaginary parts. In that case Eq. (\ref{12})
is not valid, and  
this results in four equations for the
two
unknowns $\Delta_1$ and $\Delta_2$.  These four equations
are consistent only if $\Delta_1 =0$ or $\Delta_2 =0$, which means that
there could not be any mixing between the two components. So
mixed-symmetry states are allowed only for mixing angles $\theta = 0,
\pi/2, \pi, $ and $3\pi/2$, or for $C = \pm 1, $ and  $\pm i$ and we shall
consider only these cases in the following. Essentially the same conclusion was reached
in a different context in 
Ref. \cite{cp1} in a discussion of mixed-symmetry  s-d states on the continuum. In that
case mixing was also possible only for mixing angles $0, \pi/2, \pi$ and $3\pi/2$.
However, in Ref. \cite{cp1} the discrete sum over ${\bf q}$ was replaced by a continuum
integral over angle.  
We note that  the $C=\pm i$ cases
are physically equivalent and lead to the same gap equations
(\ref{10}) and (\ref{11}). On the other hand, $C=\pm 1$ cases
corresponding to $\theta =0$ and $\pi$, respectively, lead to two
different sets of gap equations (\ref{7a}) and (\ref{8a}).

The ultraviolet momentum-space divergence of the Bardeen-Cooper-Schrieffer
 equation was
originally neutralized by a physically-motivated Debye cut off
\cite{e1,e2}. 
This procedure had the advantage of reproducing the experimentally
observed isotope
effect. 
It can also be handled by using the technique of renormalization
\cite{5a1,5a2,5a3}.  
Here we introduce a cut off in the momentum sums of the gap
equation. As there is no
pronounced isotope effect in the high-$T_c$ cuprates, the present cut off
is merely a mathematical one without any reference to the phonon-induced
Debye cut off.
   In Eqs. (\ref{7})  and (\ref{8})
both the interactions $V_1$ and
$V_2$ are assumed to be energy-independent constants for $|\epsilon_{\bf
q} - \mu| < k_B T_D$ and zero for $|\epsilon_{\bf q} - \mu| > k_B T_D$,
where $k_B T_D$ is the present  cut off.

\section{Numerical Result}

We solve the coupled set of equations (\ref{7a}) and (\ref{8a})  or
(\ref{10})  and (\ref{11})  in conjunction with the number equation
(\ref{5}). This model gives us the opportunity to study the
mixed-symmetry state composed of $d_{x^2-y^2}$ and $d_{xy}$ for different
filling and temperature for both tetragonal and  orthorhombic lattice.
This study
will also provide the interesting
variation of the chemical potential and critical temperature with
filling. We solve the above sets of equations numerically by the
method of iteration and calculate the gaps $\Delta_1$ and $\Delta_2$ at
various fillings and temperatures. 

We perform calculations on a perfect square lattice with
second-nearest-neighbor hopping contribution. In addition to the
superconductivity in coupled $d_{x^2-y^2}+ d_{xy}$ and
$d_{x^2-y^2}+id_{xy}$ waves we also consider the uncoupled $d_{x^2-y^2}$
and $d_{xy}$ waves.  The effect of orthorhombic distortion on our
investigation is also studied.  The results of our study have interesting
variation as the second nearest hopping parameter is varied and this is
studied in detail in the following for mixing angles 0 and $\pi/2$.
Throughout the calculation we consider the cut off $k_B T_D=0.1t $ eV with
the parameter $t=0.2586 $ eV.  This corresponds to a cut-off of $T_D =
300$ K. The results for the chemical potential $\mu$ and order parameters 
$\Delta_{x^2-y^2}$  and $\Delta_{xy}$ presented in this work are all in
units of $t$.

First, we study the superconductivity in uncoupled $d_{x^2-y^2}$ and
$d_{xy}$ waves by setting $\Delta_2 =0$ and $\Delta_1 =0$, respectively,
in the equations above at $T=0$ on square lattice ($\beta =1$).  For
$d_{x^2-y^2}$ wave we use potential
parameters $V_1 = 0.73t$ and $V_2=0$, and for $d_{xy}$ wave we use $V_1=0$
and $V_2=2.1t$.  The resulting $\Delta$'s with the variation of filling
$n$ are plotted in Figs. 1(a) and 1(b), respectively, for three different
values of the second-nearest-neighbor hopping parameter $\gamma$: $\gamma
=$ 0, 0.1 and 0.2. 	In the present study we shall always use 
$V_1=0.73t$ on square lattice; for $\gamma=0$ this leads to a critical
temperature of 71 K \cite{13a2}. 
The value of $V_2$ would be chosen to
have a reasonable mixture between the two components. 

From Fig. 1(a) we find that for $\gamma=0$,
$\Delta_1\equiv\Delta_{x^2-y^2}$ has a pronounced maximum at half filling 
and it
reduces quickly to zero as $n$ decreases. As second nearest hopping
$\gamma$ increases, the position of this maximum shifts to smaller $n$.
For $\gamma =0.2$, this maximum is at $n\approx 0.8$. In Fig. 1(b), we see
that for $\gamma=0$, $\Delta_2 \equiv \Delta_{xy}$ also has a maximum at
$n=1$;  however, this maximum is much wider than that in the case of
$\Delta_1$.  The position of this maximum also shifts towards smaller $n$
as the parameter $\gamma$ increases.  However, the change in the nature of
the curves as $\gamma$ increases is more pronounced for $d_{x^2-y^2}$ in
Fig. 1(a) than for $d_{xy}$ in Fig.  1(b). The variation of $\Delta $ with
$n$ in Fig. 1(a) is also quite distinct from that in Fig. 1(b). 

Now we study superconductivity in coupled $d_{x^2-y^2}+d_{xy}$ wave at
$T=0$ on a square lattice, which corresponds to  mixing angle 0. 
In this case Eqs. (\ref{7a}) and (\ref{8a}) 
are applicable. It is found that with $\gamma=0, \beta=1, $ and
$V_1=0.73t$, mixing between the two components occur for $V_2 > 1.7t$. 
 Here we
employ a 
$V_2$ slightly larger than this critical value to have adequate mixing.
The
parameters for this
model are the following: $\beta=1$, $V_{1}=0.73t$, and $V_2=2.1t$.  In
Fig. 2 we plot $\Delta_1\equiv\Delta_{x^2-y^2}$ and
$\Delta_2\equiv\Delta_{xy}$ in this case for different filling $n$ and for 
$\gamma =0$, 0.05, 0.1, and 0.2.  Although the nature of $\Delta_1$ of
Fig. 2 is similar to that of  $\Delta_1$ of Fig. 1(a), the present 
maxima are narrower. The order parameter $\Delta _2$ in the present
mixed-symmetry case has a different nature for larger $n$ compared to 
the uncoupled case. Here, in the presence of second-neighbor hopping, as
$n$ decreases, $\Delta_2$ first decreases rapidly
to a minimum  and then increases and finally decreases to zero.    
Moreover, the order parameter   $\Delta_2$ for this mixed
wave has a minimum at the same position of the maximum of $\Delta_1$.

Next we study superconductivity in coupled $d_{x^2-y^2}+id_{xy}$ wave at
$T=0$ on a square lattice, governed by Eqs. (\ref{10}) and (\ref{11}),
which
corresponds to the mixing angle $\pi/2$
or $3\pi/2$.  
The parameters for this model are the following:  $\beta=1$, 
$V_{1}=0.73t$, and $V_2=1.7t$. In this case to have mixing between the
two components for $\gamma=0$ the critical value of $V_2$ is $1.0t$. 
 In Fig. 3 we plot $\Delta_1\equiv\Delta_{x^2-y^2}$ and
$\Delta_2\equiv\Delta_{xy}$ in this case for different filling $n$ and 
$\gamma =0$, 0.05, 0.1, and 0.2.  The qualitative nature of $\Delta_1$ of
Fig. 3 is similar to that of Figs. 1(a) and 2.  The order parameter 
$\Delta_2$
for this coupled wave also has a minimum at the same position of the
maximum of $\Delta_1$. The minimum in $\Delta_{xy}$ is less pronounced in
this case than in the case of $d_{x^2-y^2}+d_{xy}$ considered in Fig. 2.

Next we study the behavior of the mixed-symmetry states in the presence of
orthorhombic distortion ($\beta=0.95$). First we consider the
$d_{x^2-y^2}+d_{xy}$ wave. In this case we needed to increase $V_1$ to
$0.97t$ in
order to have a  critical temperature of  71 K for $\gamma=0$ \cite{13a2}. 
The parameters for this model are the following: $\beta =0.95$, 
$V_1=0.97t$ and $V_2=2.1t$. To have mixing in this case the critical value
of $V_2$ is $1.5t$ for $\gamma=0$. 
In Fig. 4 we plot
$\Delta_1$ and $\Delta_2$ in this
case for different filling $n$ and  second neighbor parameter 
$\gamma =0$, 0.05, 0.1, and 0.2. We
find significant change in the behavior of the order parameters in this
case compared to the case of square lattice discussed above. In this case
$\Delta_{xy}$ could become negative for non-zero values of the
second-neighbor hopping parameter $\gamma$ for values of $n$ close to
half filling  and  the phase of the mixing changes with a
variation of
$n$ in the presence of second-nearest hopping $\gamma$.  For a non-zero 
$\gamma$ the mixing angle
$\theta$ changes from 0 to $\pi$ as $n$ increases. Also, with the increase
of
$\gamma$ the zero of $\Delta_{xy}$ shifts to a lower value of $n$,

Finally, we consider the 
$d_{x^2-y^2}+id_{xy}$ wave on orthorhombic lattice with the following
parameters:
 $\beta =0.95$, $V_1=0.97t$ and $V_2=1.7t$. For $\gamma =0$, the critical
$V_2$ for mixing
in this case is $1.2t$.
In Fig. 5 we plot
$\Delta_1$ and
$\Delta_2$ in this case for different filling $n$ for
$\gamma =0$, 0.05, 0.1, and 0.2. Here, the maximum of
$\Delta_{x^2-y^2}$
is also accompanied by the minimum of $\Delta_{xy}$. The order parameter 
$\Delta_1$ behaves in a  similar way as in Fig. 4, but $\Delta_2$
does not become negative. Both in Figs. 4 and 5 we observe a dip in
$\Delta_1$ near the maximum. This behavior was not found in Figs. 2 and 3
for square lattice.  In Figs. 2 $-$ 5 we do not find any superconductivity
for very low values of filling.

Now we investigate the temperature dependence of the order parameters in
different cases. We studied several cases for different values of $n$,
$\gamma$ and $\beta( = 1, 0.95)$. The qualitative nature of the
temperature dependence in different cases is quite similar, e. g., the
order parameters start from a large non-zero value at $T=0$. With the
increase of temperature they first remain essentially constant and 
then decrease very slowly. With further increase of temperature they
decrease more rapidly and eventually become zero at a
large enough temperature. Figure 6 illustrates a typical example of this
behavior for $\beta=1$, $n=0.8$, and $\gamma=0.2$ for the
$d_{x^2-y^2}+d_{xy}$ and $d_{x^2-y^2}+id_{xy}$ cases. The potential
parameters in these cases are the same as in Figs. 2 and 3. The present 
temperature dependence of the order parameters is quite similar to our
earlier studies for $n=1$ for $d_{x^2-y^2}+d_{xy}$ and
$d_{x^2-y^2}+id_{xy}$ waves \cite{13a1,13a2}.

In Fig. 7 we plot the chemical potential $\mu$ versus $n$ for the four
different models considered in Figs. 2 $-$ 5 for the $d_{x^2-y^2}+d_{xy}$
and $d_{x^2-y^2}+id_{xy}$ cases on tetragonal  and 
orthorhombic lattice. The parameters for the model are the same as in
Figs. 2 $-$ 5. Surprisingly enough we find that $\mu$ is essentially the
same for the two types of mixing $d_{x^2-y^2}+d_{xy}$ and
$d_{x^2-y^2}+id_{xy}$,  independent of the lattice symmetry. However,
$\mu$
depends on the parameter $\gamma$ denoting second-neighbor hopping. So in 
Fig. 7 we have four different curves for four different values of  
second-neighbor hopping. The nature of variation of $\mu$ with
$n$ changes with $\gamma$. The fall in $\mu$ with $n$ is most rapid in the
case of 
$\gamma =0$ and slows down as we increase $\gamma$.

In Figs. 8 (a) and 8 (b)  we plot $T_c$ versus $n$ for the models of Figs.
2 and 3 for three values of the second-neighbor hopping parameter
$\gamma$: 
0, 0.1, and 0.2.  We find that  $T_c$  has a  pronounced maxima as a
function of $n$. The maximum value of $T_c$
increases with the increase of $\gamma$ for both 
$d_{x^2-y^2}+d_{xy}$ and $d_{x^2-y^2}+id_{xy}$ waves. A similar increase
in the superconducting critical temperature with second-nearest-neighbor
hopping term was also
observed in a recent study of temperature and filling dependence of the
superconducting phase in the Penson-Kolb-Hubbard model\cite{cm}.

Finally, we study the interesting ratio $2\Delta_{\mbox {max}}/T_c$
as function of filling $n$ in different cases. The maximum value
$\Delta_{\mbox {max}} $ of $\Delta_{\bf k}$ was found numerically in the 
${\bf k}$ plane and the ratio $2\Delta_{\mbox {max}}/T_c$ was calculated
in different cases as a function of filling $n$. As a function of $n$, the
quantity
$\Delta_{\mbox {max}} $ has   a rapid variation  (not shown explicitly
here) similar to $T_c$ of Fig.
8. 
Although $T_c$ as well
as $\Delta_{\mbox {max}}$  are found to exhibit  rapid variation as
function of $n$, 
the ratio
$2\Delta_{\mbox {max}}/T_c$ is found to be a reasonably smooth function of
$n$. This ratio varies from 4.0 for $n=1$, to a maximum of about 4.7 as
$n$ decreases; the maximum is obtained for $n\approx 0.4$. 
The
average value of $2\Delta_{\mbox {max}}/T_c$ is found to be about 4.3.

\section{Summary}

In this work we have studied different properties of the mixed
$d_{x^2-y^2}+d_{xy}$ and $d_{x^2-y^2}+id_{xy}$ superconducting states in
two-dimensional cuprates using a tight-binding model with appropriate
lattice symmetry.
Recent experiments and other theoretical investigations  suggested the
formation of such mixed-symmetry states in cuprates on tetragonal and
orthorhombic lattice. We have included the
effect of orthorhombic distortion of the lattice as well as of second
nearest hopping on the order parameter of these states in the
tight-binding model. The allowed mixing angles are 0, $\pi/2, \pi,$ and 
$3\pi/2$.
We studied the variation of order parameters
$\Delta_{x^2-y^2}$ and $\Delta_{xy}$ with filling $n$ for tetragonal and
orthorhombic lattices for different second-neighbor hopping for pure and 
mixed-symmetry states. These results are reported in Figs. 1 $-$ 5. In the
case of mixed-symmetry $d_{x^2-y^2}+d_{xy}$ wave on orthorhombic 
lattice in the presence of second-neighbor hopping, the mixing angle 
  changes from 0 to $\pi$ as filling $n$ is varied. 
 We have also
studied the temperature dependence of the order parameter
under different situations. This temperature dependence is quite similar
to the $n=1$ case studied  by us previously \cite{13a1,13a2}. We also
studied
the
dependence of the 
critical temperature $T_c$ for   the mixed-symmetry state
on  filling $n$ under different situations and find that this dependence 
exhibits a maximum and rapid variation.  
The maximum of 
$T_c$ increases with the increase of second-nearest hopping. The ratio
$2\Delta_{\mbox{max}}/T_c$ is a  slowly varying function of $n$
and varies from 4 for $n=1$ to a maximum of about
4.7 as $n$ decreases.

We thank Conselho
Nacional de Desenvolvimento Cient\'{\i}fico e Tecnol\'ogico and Funda\c c\~ao
de Amparo \`a Pesquisa do Estado de S\~ao Paulo for financial support.


{\bf Figure Captions:}
\vskip 0.8cm

1. The order parameters (a) $\Delta_{x^2-y^2}$ and (b) 
$\Delta_{xy}$  at different filling $n$ for
second nearest hopping 
$\gamma = 0$ (full line), 0.1 (dashed-dotted line), and 0.2   (dashed
line) for square lattice.  The potential parameters for (a) are
$V_1=0.73t$, $V_2=0$, and for (b) are $V_1=0$, $V_2=2.1t$.

2. The  order parameters $\Delta_{x^2-y^2}$ (dashed line),
$\Delta_{xy}$ (full line)  for different filling $n$ for mixed 
 $d_{x^2-y^2}+d_{xy}$ wave  for second nearest hopping
$\gamma = 0$, 0.05, 0.1, and 0.2   on  square lattice.  The 
parameters for the model are $\beta =1$,  
$V_1=0.73t$, $V_2=2.1t$.

3. Same as in Fig. 2 for  mixed
 $d_{x^2-y^2}+id_{xy}$ wave.   The
parameters for the model are $\beta =1$,
$V_1=0.73t$, $V_2=1.7t$.

4. Same as in Fig. 2 on  orthorhombic lattice. The
parameters for the model are $\beta =0.95$,
$V_1=0.97t$, $V_2=2.1t$.

5. Same as in Fig. 3 on orthorhombic lattice. The
parameters for the model are $\beta =0.95$,
$V_1=0.97t$, $V_2=1.7t$.

6. Temperature dependence of the order parameters $\Delta_{x^2-y^2}$
and $\Delta_{xy}$ for mixed-symmetry $d_{x^2-y^2}+d_{xy}$ (dashed line)
and
$d_{x^2-y^2}+id_{xy}$ (full line)
waves for $\beta=1$, $n=0.8$, and $\gamma=0.2$. The larger order
parameters represent the  $d_{x^2-y^2}$ wave. 
The parameters of the models are the same as in Figs. 2 and 3,
respectively.

7. Chemical potential $\mu$ versus filling $n$ for the four models of 
Figs. 2 $-$ 5 for $\gamma = 0$ (full line), 0.05 (short dashed line),
0.1 (long dashed line), and 0.2 (dashed-dotted line). Four different
models lead to identical result.

8. Critical temperature $T_c$ versus $n$ for the models of (a) Fig. 2
and (b) Fig. 3 for $\gamma =0 $ (full line),  0.1 (short dashed line), and
0.2 (long dashed line).

\begin{references}

 
\bibitem{1}J. G. Bednorz,  K. A. M\"uller, Z. Phys. B { 64} 
(1986) 1898.



\bibitem{2a}H. Ding, Nature { 382} (1996) 51. \bibitem{2b}
D. J. Scalapino, Phys. 
Rep.  { 250} (1995) 329. 




\bibitem{3a} W. N. Hardy, D. A. Bonn, D. C. Morgan, R. X. Liang,  K.
Zhang,  Phys. Rev. Lett. { 70} (1993) 3999
\bibitem{3b} K. A. Moler, D. J. Baar, J. S. Urbach, R. X. Liang, W. N.
Hardy, A.
Kapitulnik, { Phys. Rev. Lett.} { 73} (1994) 2744.\bibitem{3c}
K. Gofron, J. C. Campuzano, A. A.  Abrikosov, M. Lindroos, A. Bensil,
H. Ding, D. Koelling,  
B. Dabrowski,
{ Phys. Rev. Lett.} { 73} (1994) 3302. 


\bibitem{4a} M. Prohammer, A. Perez-Gonzalez,  J. P. Carbotte, Phys.
Rev. B
{ 47} (1993) 15152. \bibitem{4b} J. Annett, N. Goldenfeld,  S.
R.
Renn,
{
Phys. Rev. B} { 43} (1991) 2778. \bibitem{4c}  
N. Momono, M. Ido, Physica C { 264} (1996) 311. \bibitem{4d}
A. J. Millis, H. Monien,  D. Pines, Phys. Rev. B { 42}
(1990) 167. \bibitem{4e}  
M. Houssa,  M. Ausloos, {Physica C} { 265} (1996) 258. 

\bibitem{5a1} 
S.  K. Adhikari, A. Ghosh, Phys. Rev. B  { 55} (1997) 1110.

\bibitem{5a2}
S.  K. Adhikari, A. Ghosh J.
Phys.:
Cond. Mat. { 10} (1998) 135.\bibitem{5a3}
 A. Ghosh, S. K. Adhikari, Eur. Phys. J
B { 2} (1998) 31. 



\bibitem{5a}A. G. Sun, D. A. Gajewski, M. B. Maple,  R. C. Dynes, Phys.
Rev. Lett. { 72} (1994) 2267.\bibitem{5b}
A. G. Sun, S. H. Han, A. S. Katz, D. A. Gajewski, M. B. Maple,   R. C.
Dynes, Phys. Rev. B { 52} (1995) R15731. \bibitem{5c}
P. Chaudhari,  S. Y. Lin, Phys.
Rev. Lett. { 72} (1994) 1084.\bibitem{5d}
M. Covington, G. Aprili, E. Paraoanu, L. H. Greene, F. Xu, J.
Zhu,  C. A. Mirkin, Phys. Rev. Lett. { 79} (1997) 277.





\bibitem{6}
K. A. Kouznetsov, A. G. Sun, B. Chen, A. S. Katz, S. R. Bahcall, J.
Clarke, R. C. Dynes, D. A. Gajewski, S. H. Han, M. B. Maple, J.
Giapintzakis, J. T. Kim,  D. M. Ginsberg, Phys. Rev. Lett. { 79}
 (1997) 3050.

\bibitem{7}
S. Sridhar, H. Srikanth, 
Z. Zhai, B. A. Willemsen, T.  Jacobs, A. Erb, E. Walker, 
R. Flukiger, Physica C {   282} (1997) 256. 

\bibitem{8a}
J. Ma, C. Quitmann, R. J. Kelley, H. Berger, G. Margaritondo,  M.
Onellion,  Science { 267} (1995) 862.\bibitem{8b}
R. J.
Kelley, C. Quitmann, M. Onellion, H. Berger, P. Almeras,  G.
Margaritindo, Science { 271} (1996) 1255.


\bibitem{9}
K. Krishana, N. P. Ong, Q. Li, G. D. Gu,  N. Koshizuka,
Science
{ 277} (1997) 83.

\bibitem{10}R. Movshovich, M. A. Hubbard, M. B. Salamon, A. V. Balatsky,
R.
Yoshizaki,
J. L. Sarrao,  M. Jaime, Phys. Rev. Lett. { 80} (1998) 1968. 



\bibitem{11a}R. B. Laughlin, Phys. Rev. Lett. { 80} (1998) 5188. 
\bibitem{11b}
M. Franz,  Z. Te\'sanovi\'c, Phys. Rev. Lett. { 80} (1998) 4763.
\bibitem{11c} M. I.  Salkola, J. R. Schrieffer, Phys. Rev. B { 58}
(1998) R5952.


\bibitem{131} M. Sigrist, D. B. Bailey, R. B. Laughlin, Phys. Rev.
Lett.
{ 74} (1995) 3249.\bibitem{132}  M. Liu, H. Y. Teng, D. Y. Xing, 
J. M. Dong,
Physica C { 282-287} (1997) 1649.\bibitem{133}  C. O'Donovan,  J.
P.
Carbotte,
Phys.
Rev. B { 52} (1995) 16208. \bibitem{134} C. O'Donovan,  J. P.
Carbotte, Physica
C { 252}
(1995) 87. \bibitem{135}
A. Ghosh,
 S. K. Adhikari, J. Phys.:  Cond. Mat. { 10}
(1998) L319.\bibitem{136} A. Ghosh,  S. K. Adhikari,
  Physica C { 322} (1999) 37.

\bibitem{13a1}A. Ghosh,  S. K. Adhikari,
Physica C
{ 309} (1998) 251.
\bibitem{13a2}A. Ghosh,  S. K. Adhikari,    Phys. Rev. B {
60} (1999) 10401.


\bibitem{14a}J. H. Xu, J. L. Shen, J. H. Miller, Jr., 
 C. S. Ting, Phys. Rev. Lett. { 73} (1994) 2492.\bibitem{14b}
M. Matsumoto,  H. Shiba, 
J. Phys. Soc. Japan { 64} (1995) 3384.\bibitem{14c}
E. A. Shapoval, JETP Lett. {64} (1996) 625.\bibitem{14d}
A. E. Ruckenstein, P. J. Hirschfeld,  J. Apel, Phys. Rev. B
{36} (1987) 857.
\bibitem{14e} G. Kotliar, { Phys. Rev. B} {37} (1988) 3664.  




\bibitem{ba}
D. N. Basov, R. Liang, D. A. Bonn, W. N. Hardy, B. Dabrowski, M. Quijada,
D. B. Tanner, J. P. Price, D. M. Ginsburg,  T. Timusk, Phys. Rev. Lett. 
 74 (1995) 598.



\bibitem{cp1} K. A. Musaelian, J. Betouras, A. V. Chubukov, R. Joynt, Phys. Rev. B
53 (1996) 3598


\bibitem{e1}
J. Bardeen, L. N. Cooper, J. R. Schrieffer, Phys. Rev.  { 108}
  (1957) 1175.
\bibitem{e2}
M. Tinkham, {\it Introduction to Super\-conductivity}, (Mc\-Graw-\-Hill Inc.,
New York, 1975).


\bibitem{cm}F. Dolcini,  A. Montorsi, Phys. Rev. B 62 (2000) 2315. 






\end{references}
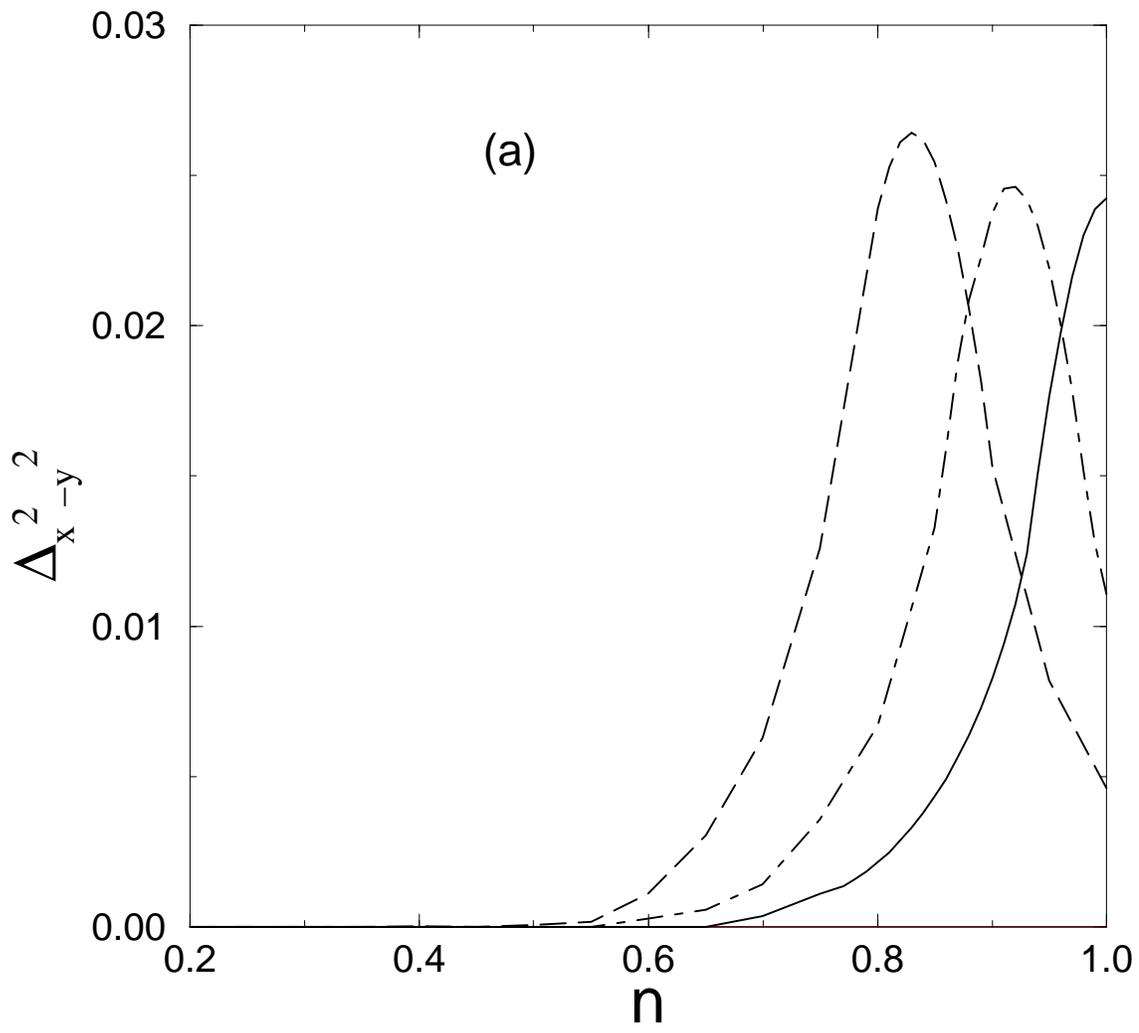
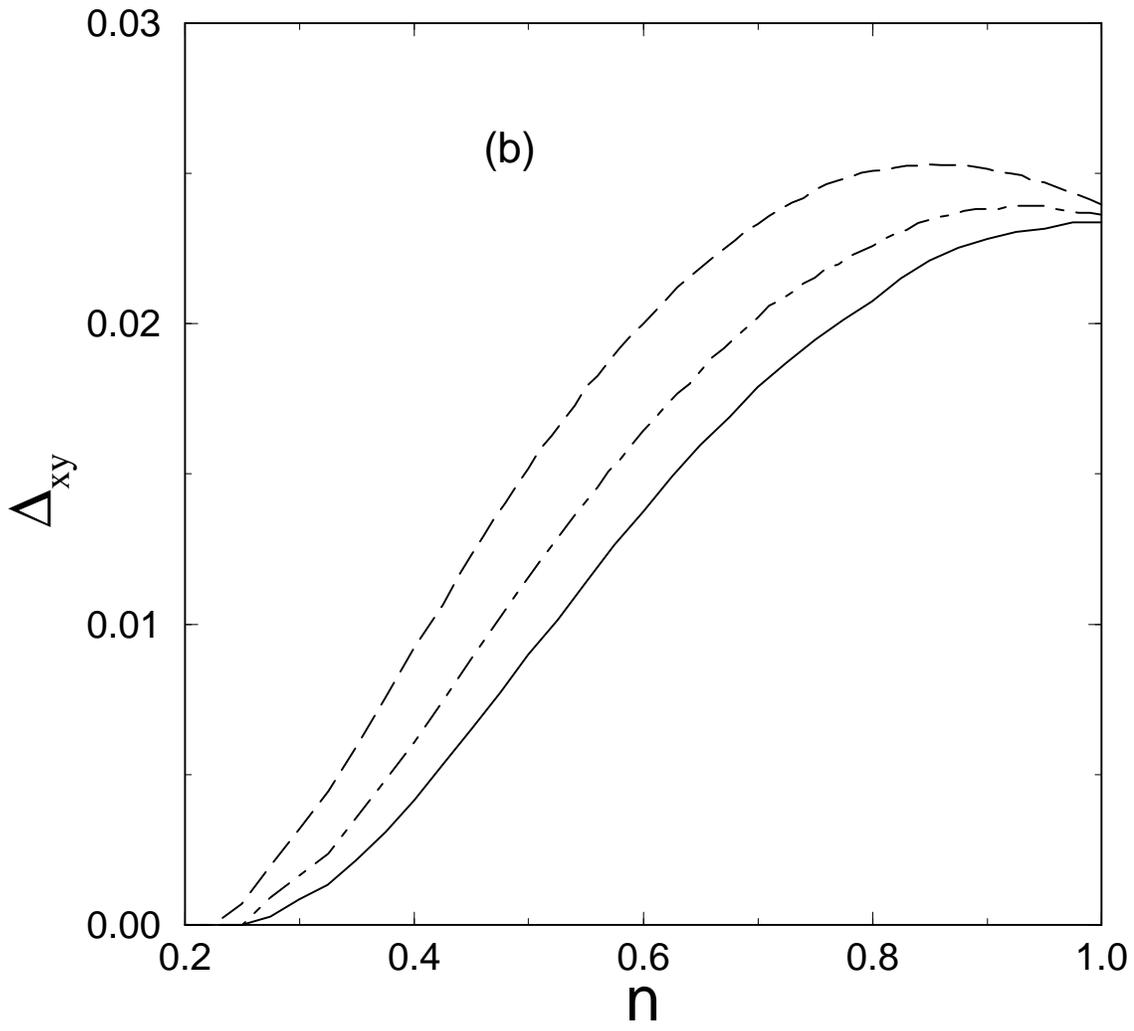
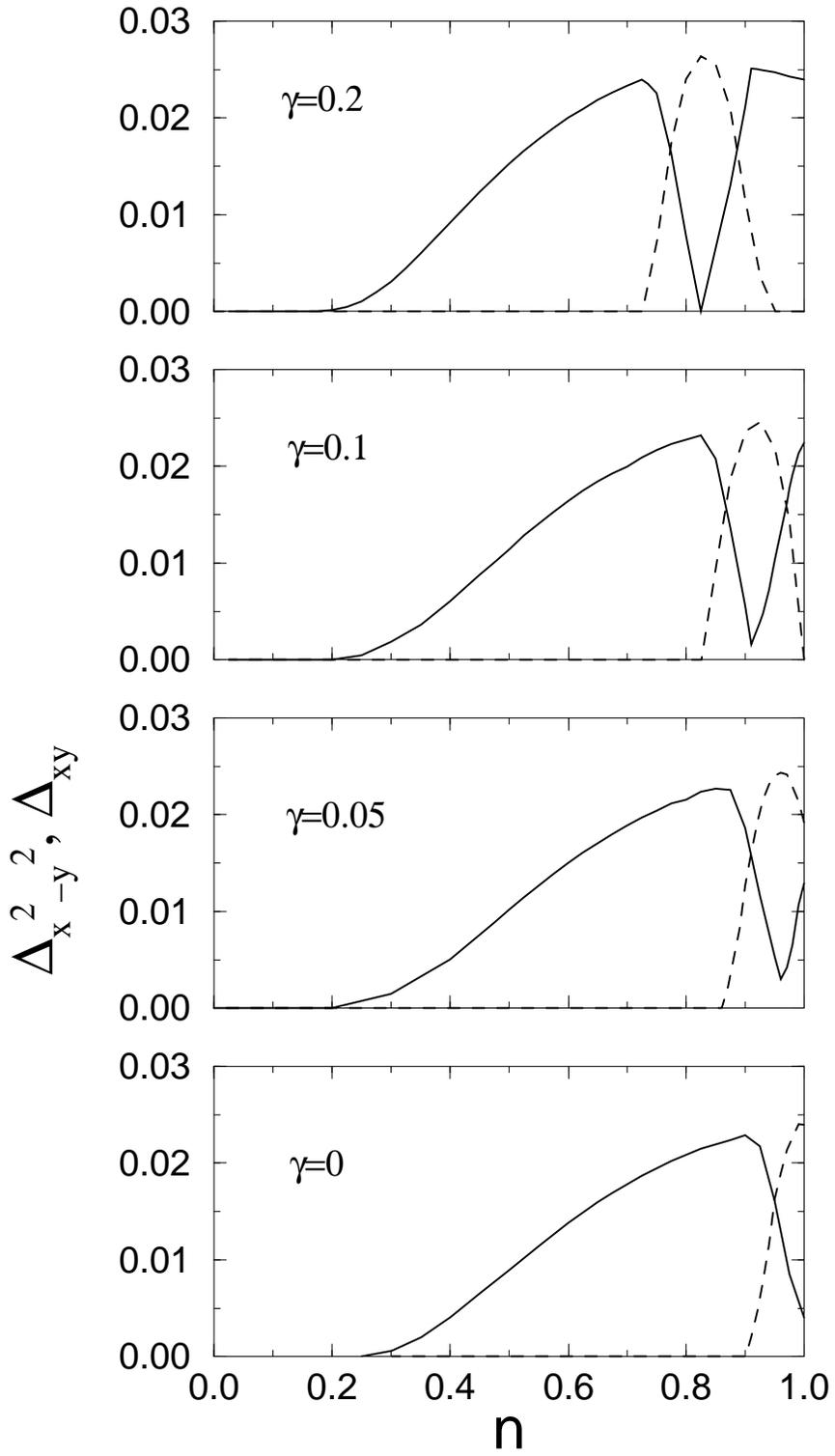
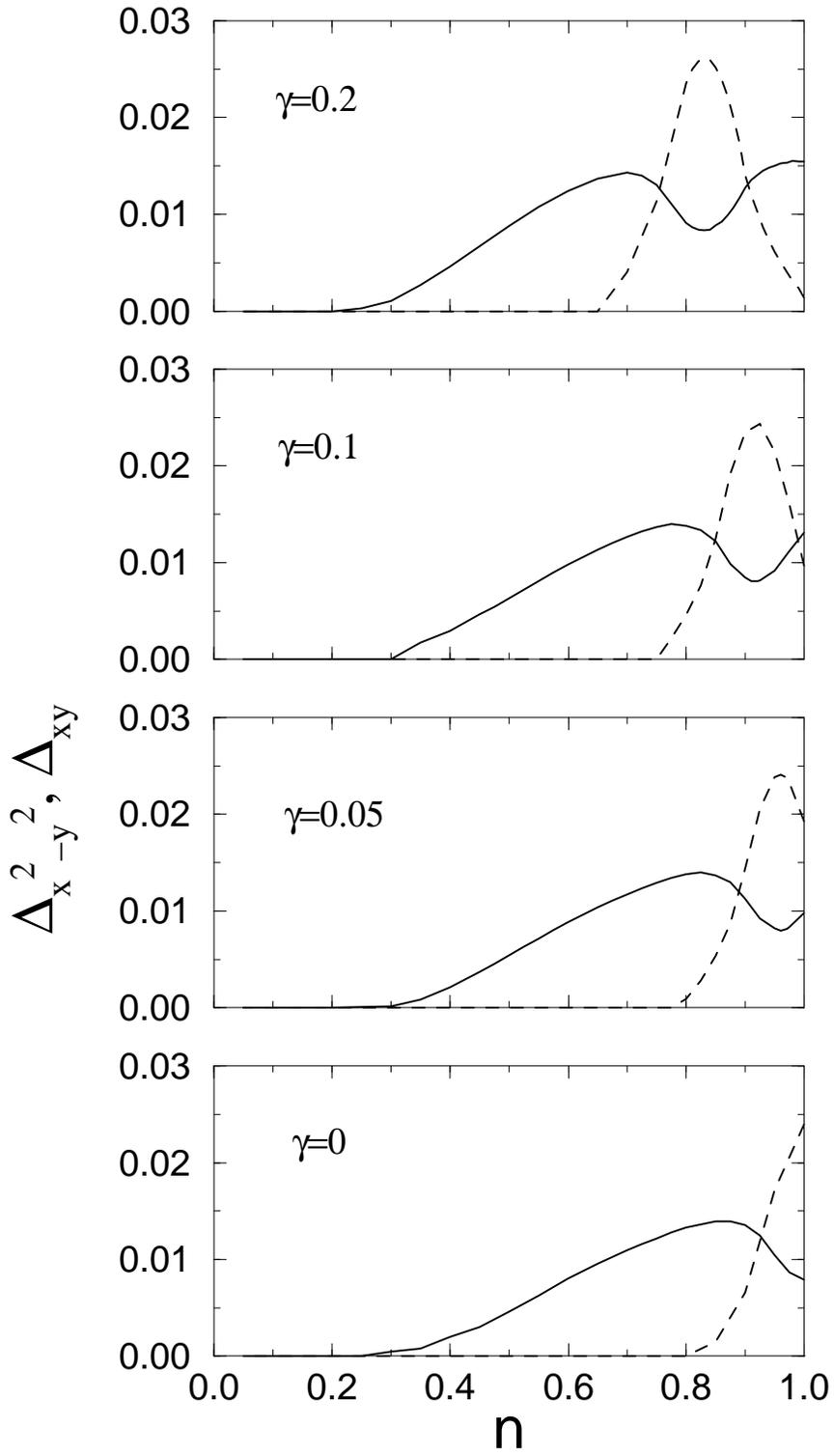
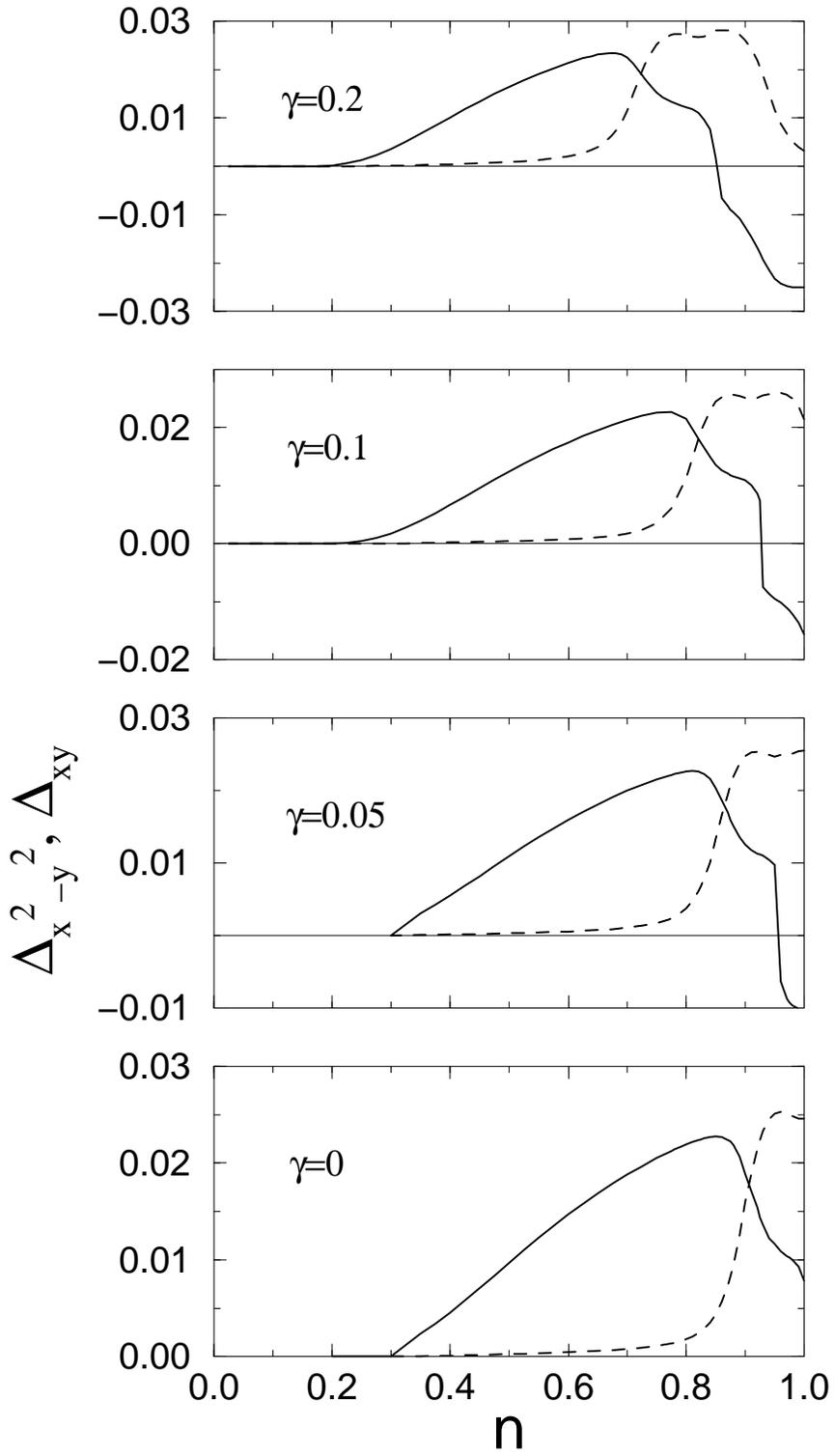
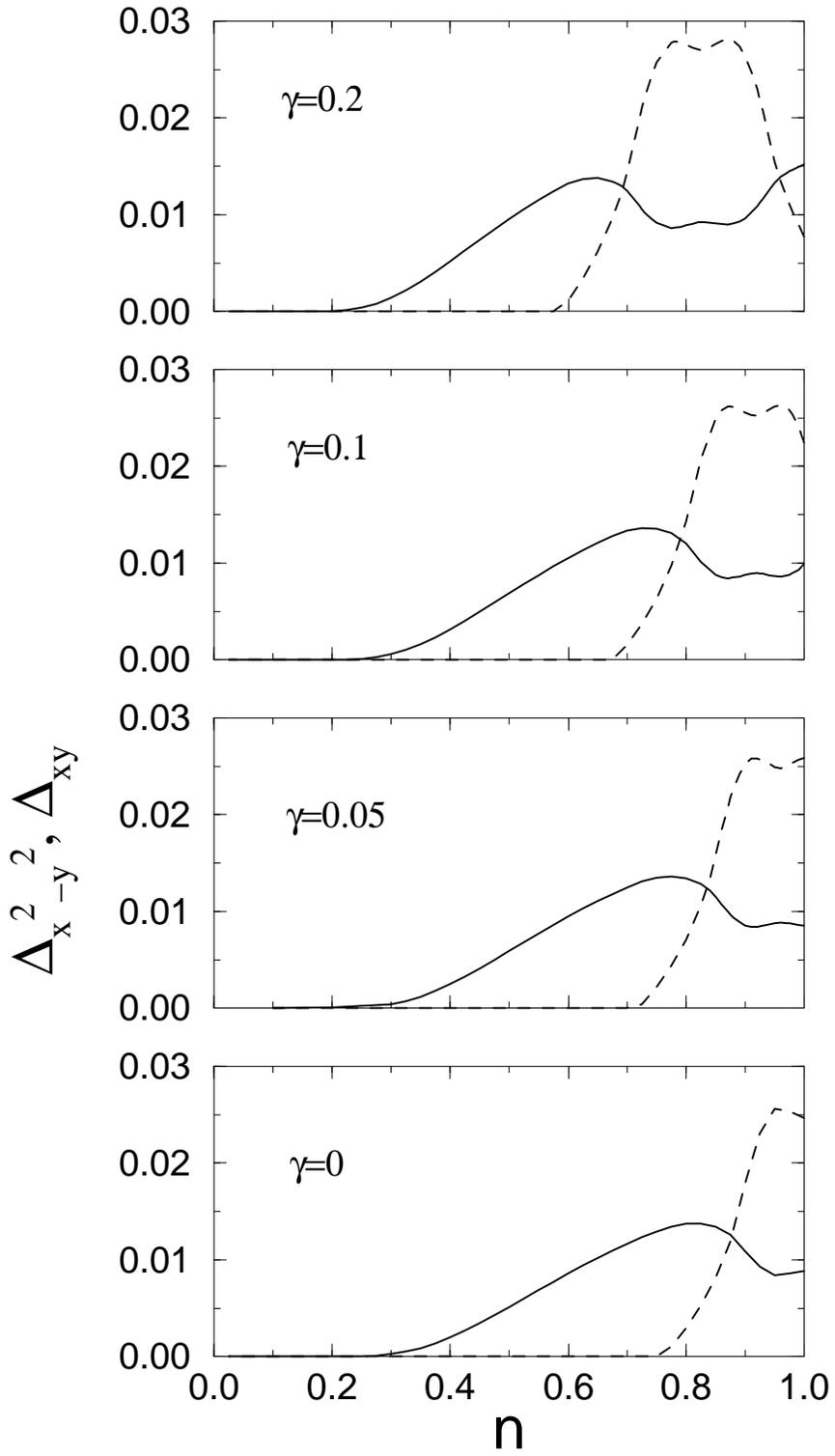
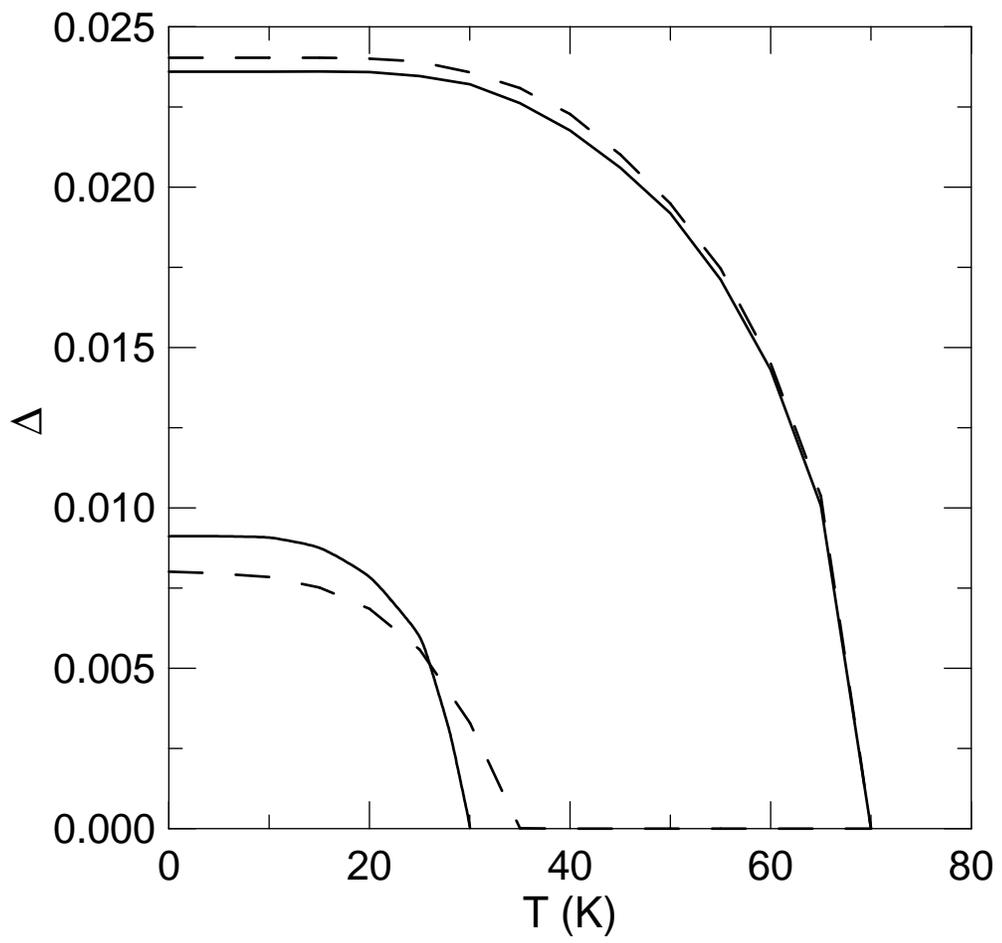
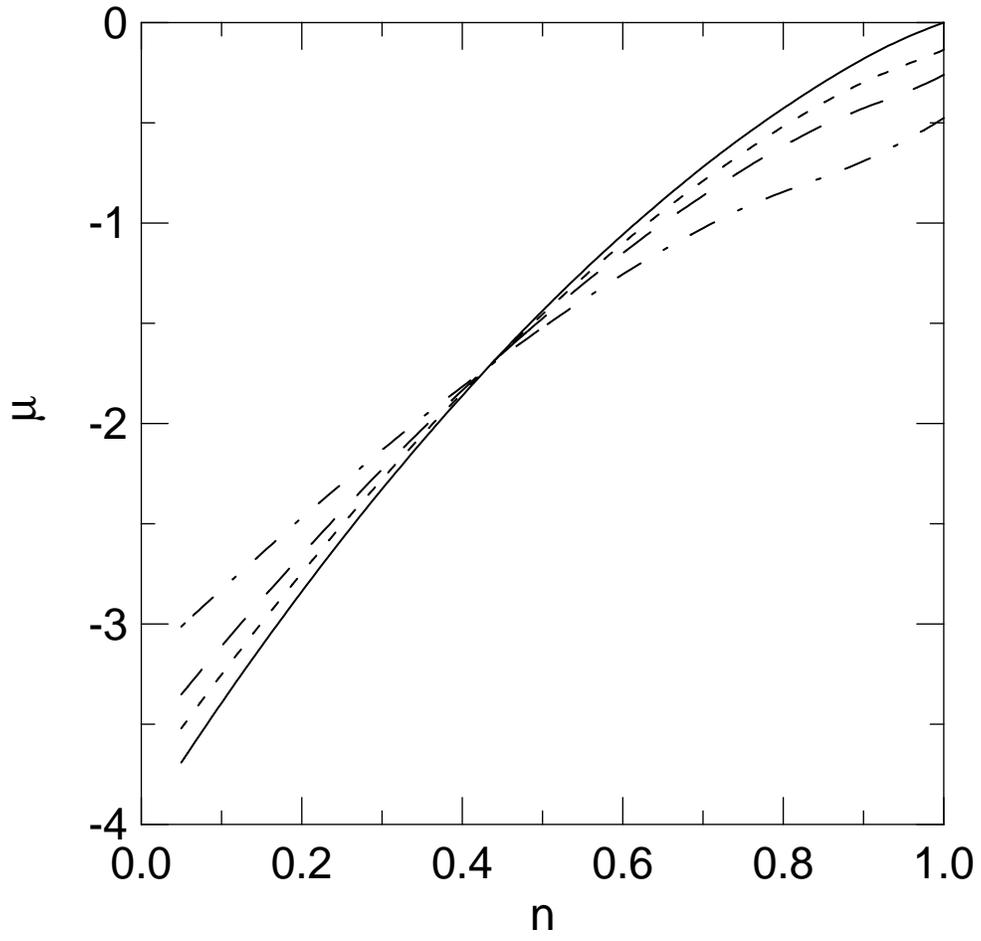
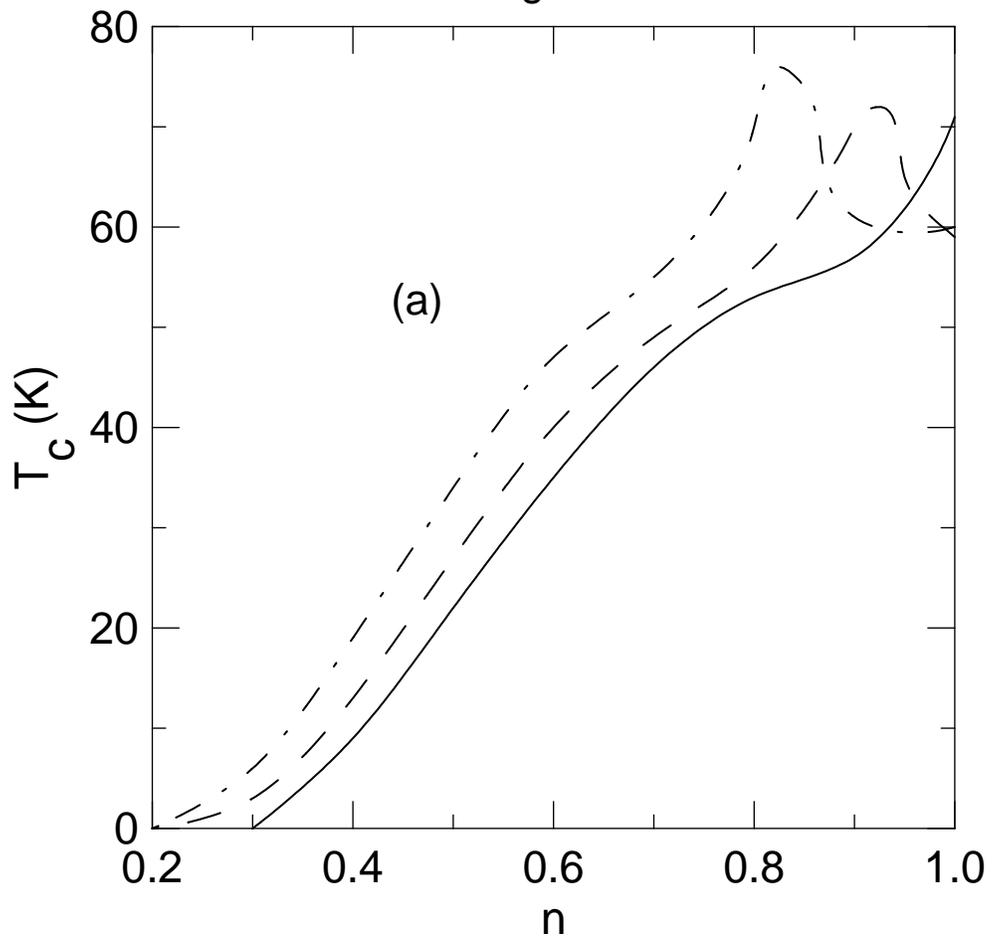
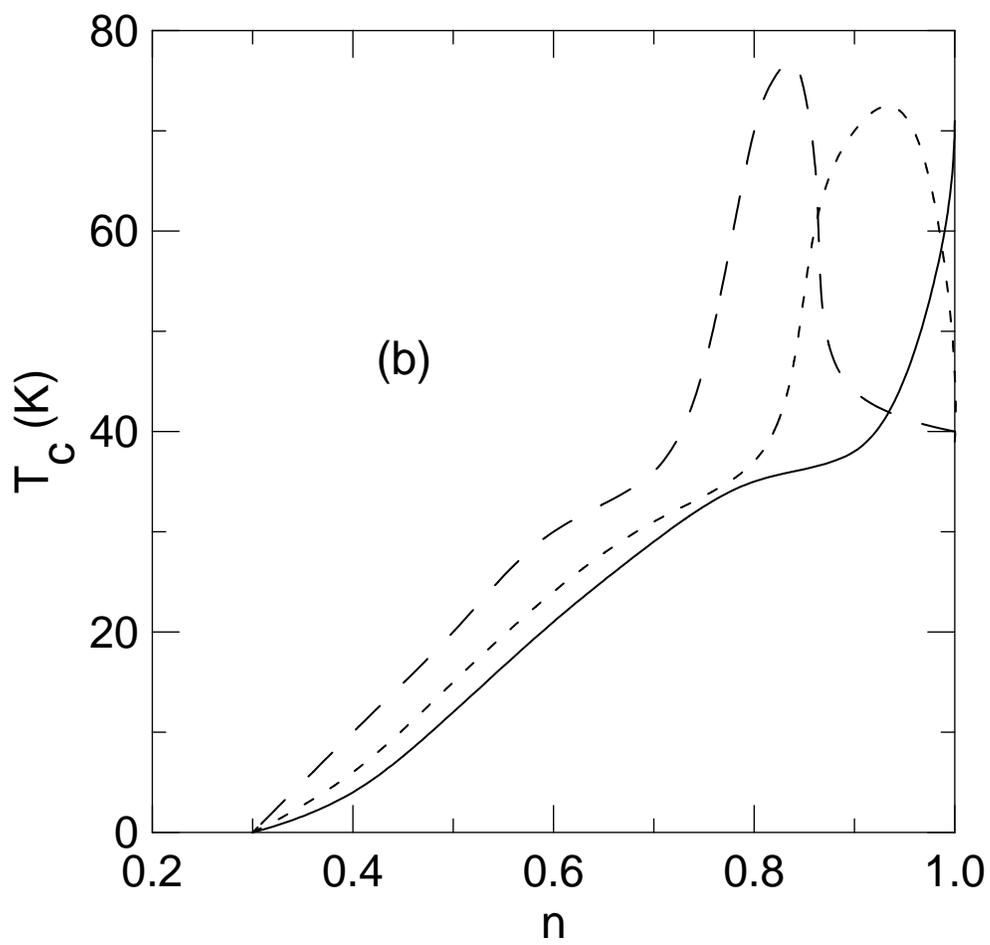
\end{document}